\documentclass[prd,preprint,showpacs,eqsecnum]{revtex4}

\usepackage{amsfonts}
\usepackage{amssymb}
\usepackage{graphicx}

\newcommand{\be}{\begin{equation}}
\newcommand{\bea}{\begin{eqnarray}}
\newcommand{\ee}{\end{equation}}
\newcommand{\eea}{\end{eqnarray}}
\newcommand{\bpi}{\begin{picture}}
\newcommand{\bce}{\begin{center}}
\newcommand{\epi}{\end{picture}}
\newcommand{\ece}{\end{center}}

\newcommand{\lslush}{k\hspace{-0.20cm}/}
\newcommand{\pslush}{p\hspace{-0.18cm}/}

\newcommand{\sw}{s_{\rm w}}
\newcommand{\cw}{c_{\rm w}}
\newcommand{\gw}{g_{\rm w}}

\newcommand{\xiw}{\xi_{\scriptscriptstyle W}}
\newcommand{\lw}{\lambda_{\scriptscriptstyle W}}
\newcommand{\etaw}{\eta_{\scriptscriptstyle W}}
\newcommand{\MSW}{M^2_{\scriptscriptstyle W}}

\newcommand{\sumv}{\sum_{V}}

\newcommand{\Mw}{M_W}

\newcommand{\PrL}{P_{\!\scriptscriptstyle{L}}}
\newcommand{\PrR}{P_{\!\scriptscriptstyle{R}}}
\newcommand{\JV}{J^e_{V_\alpha}}

\begin{document}
\begin{flushright}
FTUV-04-0528\\ 
IFIC-04-20\\
ECT*-04-06
\end{flushright}

\title{The neutrino charge radius in the presence of fermion masses}

\date{May 28, 2004}

\author{J. Bernab\'eu} 
\author{J. Papavassiliou}
\affiliation{Departamento de F\'\i sica Te\'orica$,$ and IFIC Centro Mixto, Universidad de Valencia-CSIC, E-46100, Burjassot, Valencia, Spain}
\author{D. Binosi}
\affiliation{ECT*, Villa Tambosi, Strada delle Tabarelle 286 I-38050 Villazzano (Trento), Italy, \\
and\\
I.N.F.N., Gruppo Collegato di Trento, Trento, Italy}

\begin{abstract}

We show how the crucial gauge cancellations 
leading to a physical 
definition of the  neutrino charge radius persist  
in the presence of non-vanishing fermion masses.
An explicit one-loop calculation
demonstrates 
that, as happens in the massless case,
the pinch technique rearrangement of the Feynman amplitudes, 
together with the judicious exploitation of the 
fundamental current relation  
$J_{\alpha}^{(3)} = 2(J_{Z} + \sin \theta_{\rm w}^2J_{\gamma})_{\alpha}$, 
leads to a completely gauge independent definition of 
the effective neutrino charge radius.
Using the formalism of the Nielsen identities it is further proved that 
the same cancellation mechanism operates unaltered to all orders in 
perturbation theory.

\end{abstract}


\maketitle

\section{Introduction}

It is well-known that, even though within  the Standard  Model (SM)  the  photon ($A$) 
does  not  interact with  the neutrino  ($\nu$) at  tree-level,
an   effective  photon-neutrino vertex ${\Gamma}^{\mu}_{A \nu \bar{\nu}}$ 
is  generated  through  one-loop  radiative  corrections, giving   rise   to   a   non-zero neutrino   charge   radius   (NCR) \cite{Bernstein:jp}.
Traditionally (and, of course, rather heuristically) 
the NCR has been interpreted as a measure of the ``size'' of the neutrino $\nu$ 
when probed electromagnetically, owing to its classical definition \cite{footnote}
(in the static limit) as the second moment 
of the  spatial neutrino charge density $\rho_{\nu}(\bf{r})$, {\it i.e.} 
$\big <r^2_{\nu}\,  \big> = \int\! d {\bf{r}}\, r^2 \rho_{\nu}({\bf{r}})$.
However, the direct calculation of this
quantity has been faced   with serious complications \cite{Lucio:1984mg}
which,  in turn, can be traced  back to the  fact  that in non-Abelian gauge   theories   off-shell  Green's functions depend in general
explicitly on   the   gauge-fixing parameter.
 In the   popular renormalizable
($R_{\xi}$) gauges, for example, the electromagnetic
form-factor $F_1$ depends  explicitly  on gauge-fixing parameter $\xi$
 in a
prohibiting  way. Specifically, even though  in  the static limit of
zero momentum transfer,  $q^2 \to 0$,  the form-factor $F (q^2,\xi)$
becomes independent of  $\xi$,  its first  derivative with respect  to
$q^2$, which corresponds to the definition  of the NCR, namely
$\big <r^2_\nu\,  \big>    = -\, 6\, (d F /dq^2)_{q^2=0}$\, , 
continues to depend on it.
Similar (and some times worse) problems occur in the context of other
gauges ({\it e.g.} the unitary gauge).

One way out of this difficulty is to 
identify  a {\it  modified} vertex-like  amplitude, which
could  lead to  a consistent  definition of  the  electromagnetic form
factor and the corresponding NCR. The basic idea is to 
exploit  the fact that the  {\it full} one-loop
$S$-matrix element describing the  interaction between a neutrino with
a  charged particle  is gauge-independent,  and try  to  rearrange the
Feynman graphs contributing to this scattering amplitude in such a way
as to find a vertex-like  combination that would satisfy all desirable
properties.  What became  gradually clear over the years  was that, for
reaching  a   physical  definition  for   the  NCR,  in   addition  to
gauge-independence, a plethora  of important physical constraints need
be    satisfied.   For   example,   one    should   not   enforce
gauge-independence      at     the     expense      of     introducing
target-dependence. Therefore, a  definite guiding-principle is needed,
allowing  for the systematic  construction of  physical sub-amplitudes
with   definite  kinematic   structure  ({\it   i.e.},  self-energies,
vertices, boxes).

The field-theoretical methodology allowing this 
meaningful rearrangement of the perturbative expansion  
is that of the pinch technique (PT) 
\cite{Cornwall:1982zr}. The  PT  is  a
diagrammatic  method which exploits  the  underlying symmetries 
encoded  in  a  {\it  physical} amplitude,  such  as  an  $S$-matrix  element,
in  order  to  construct effective   Green's   functions    with   special  
properties.  
In the context of the NCR,
the basic observation \cite{Bernabeu:2000hf}
is that the gauge-dependent parts of the conventional 
$A^{*}\nu \nu$,  to which one would naively associate 
the (gauge-dependent) NCR, 
communicate and eventually cancel {\it algebraically}
against analogous  
contributions concealed inside the $Z^{*}\nu \nu$ vertex, the self-energy graphs,
and the box-diagrams (if there are boxes in the process), 
{\it before}   any integration over the 
virtual momenta is carried out. For example, due to rearrangements produced through the 
systematic triggering of elementary Ward identities,  
the gauge-dependent contributions 
coming from boxes 
are not box-like, but self-energy-like or vertex-like; it is only those latter
contributions that need be included in the definition of the new, effective 
$\widehat{A}^{*}\nu \nu$ vertex.

The new one-loop proper three-point function satisfies the following properties
\cite{Bernabeu:2002nw,Bernabeu:2002pd,Papavassiliou:2003rx}:
({\it i}) is independent of the 
gauge-fixing parameter ($\xi$);
({\it ii}) is ultra-violet finite;
({\it iii}) satisfies a QED-like Ward-identity;
({\it iv}) captures all that is coupled to a genuine $(1/q^2)$ photon propagator,
{\it before} integrating over the virtual momenta; 
({\it v}) couples electromagnetically to the target;
({\it vi}) does not depend on the $SU(2)\times U(1)$ quantum numbers
of the target-particles used;
({\it vii}) has a non-trivial dependence on the mass $m_i$ of the 
charged iso-spin partner $f_i$ of the neutrino in question; 
({\it viii}) contains only physical thresholds;
({\it ix}) satisfies unitarity and analyticity;
({\it x}) can be extracted from experiments.\\
Notice in particular that the properties from ({\it iv}) to ({\it vi})
ensure that the quantity constructed is a genuine photon vertex, 
uniquely defined in the sense that it is independent of using
either weak iso-scalar sources (coupled to the $B$-field) or weak iso-vector sources
(coupled to $W^0$), or any other charged combination.
As for property ({\it x}), the NCR defined through this procedure may
be  extracted  from  experiment,  at  least  in  principle, 
by expressing  a set  of  experimental electron-neutrino
cross-sections in  terms of the  finite NCR and two  additional gauge-
and  renormalization-group-invariant quantities, corresponding  to the
electroweak effective charge and mixing angle \cite{Bernabeu:2002nw,Bernabeu:2002pd}.

Given the progress achieved with the above properties for the NCR, 
it is important to address some of the remaining open theoretical issues. 
To begin with, 
in the construction presented in \cite{Bernabeu:2000hf}
it has been tacitly assumed that the mass of the iso-doublet partner 
of the neutrino under consideration vanishes (except when needed 
for controlling infra-red divergences). That should be a acceptable approximation,
given the fact that the fermion masses are naturally suppressed numerically, relative
to the relevant physical 
scale, {\it i.e.}, the mass of the $W$-boson, provided that the neglected 
pieces form themselves a gauge-independent sub-set. 
As has been pointed out \cite{Fujikawa:2003ww}, when 
computing the {\it conventional} vertex (that is, before applying any PT rearrangements) 
in the $R_{\xi}$ gauges and keeping the fermion masses non-zero, a gauge-dependent term
survives, which, in fact, diverges at $\xi =0$. It is therefore an indispensable exercise to verify 
that the PT procedure in the presence of masses indeed identifies precisely the 
contributions which will cancell such pathological terms, exactly as happens in the 
massless case, {\it without} any additional assumptions. 
In this article we will undertake this task and demonstrate through
an explicit calculation how  
the (vertex-like) gauge-dependent contribution proportional to the fermion masses 
cancel partially against
similar gauge-dependent contributions stemming from 
graphs containing would-be Goldstone bosons, and partially against 
vertex-like contributions concealed inside box-graphs, exactly as dictated by the
well-defined PT procedure. 

In addition, the constructions related to the NCR  
 have thus far 
been solely restricted to the one-loop level. Clearly, it is important to
demonstrate that the crucial cancellations and non-trivial rearrangements 
operating at one-loop persist and can be generalized to all orders.
In this article we demonstrate that the pertinent gauge-cancellations 
take place through precisely the same mechanism as at one-loop, by resorting to the 
powerful formalism of the Nielsen identities (NIs) \cite{Nielsen:1975fs}. 
These identities control in a concise, completely 
algebraic way, the gauge-dependences of individual Green's functions (such as the
off-shell photon-neutrino vertex ${\Gamma}^{\mu}_{A \nu \bar{\nu}}$ in question), 
and allow for an all-order demonstration of gauge-cancellations between various 
Green's functions, when the latter are combined to form ostensibly gauge-invariant 
quantities, such as $S$-matrix elements. 
In the present paper, using the corresponding NIs,  
we will show that the gauge-dependence of the 
vertex ${\Gamma}_{A_\mu \nu \bar{\nu}}$ has precisely the form needed for 
cancelling against analogous gauge-dependent {\it vertex-like} contributions from the boxes, employing
nothing more than the fundamental current relation
(see, for example, the sixth reference in ~\cite{Lucio:1984mg})
\be
J_{\alpha}^{(3)} = 2(J_{Z} + \sw^2 J_{\gamma})_{\alpha},
\label{fci}
\ee
with $\sw\equiv\sin \theta_{\rm w}$, and $J_{\alpha}^{(3)}$ the third iso-triplet current of $SU(2)_{L}$. 

The paper is organized as follows: In Section~\ref{gc} we present an explicit one-loop proof 
of the relevant gauge cancellations in the presence  of non-vanishing fermion masses. 
The upshot of this construction is to demonstrate that the fermion masses do not distort
in the least the crucial $s$-$t$ channel 
cancellations characteristic of the 
PT, and that, at the end of the cancellation procedure, a completely gauge-independent 
NCR emerges. In Section~\ref{aoc} we present a general all-order proof of the same 
gauge-cancellations, by means of the NIs. 
Finally, in Section~\ref{conc}, 
we present our conclusions.  

\section{\label{gc} Gauge cancellations in the presence of fermion masses}

In this section we go over the fundamental cancellation mechanism, and 
outline its basic ingredients. In particular, we emphasize the
topological modifications induced on Feynman diagrams due to
the presence of longitudinal momenta, the  
r\^ole of the current operator identity in implementing the gauge cancellations,
and explain qualitatively
the modifications induced when the fermion masses are turned on.  
Then, we carry out an explicit 
one-loop calculation in the presence of non-vanishing fermion masses,
and demonstrate 
the precise cancellations of massive gauge-dependent terms. 
In doing so we will show that no additional theoretical input or assumptions are needed, whatsoever.

\subsection{General considerations}

The topological modifications,
 which allow the communication between 
kinematically distinct graphs 
(enforcing eventually the cancellation of the gauge dependent pieces),   
are produced when  elementary Ward identities are triggered 
by the  virtual longitudinal momenta ($k$) inside Feynman diagrams, 
 furnishing {\it inverse propagators} \cite{Cornwall:1982zr}. 
The longitudinal momenta appearing in the $S$-matrix element of
$f\nu\to f\nu$ originate from the 
tree-level gauge-boson propagators and tri-linear gauge-boson vertices
appearing inside loops. In particular, in the $R_{\xi}$-scheme
the gauge-boson propagators have the general form  
\be
\Delta^{\mu\nu}_{i} (k) = -i \Bigg [g^{\mu\nu}-
\frac{(1-\xi_{i})k^{\mu}k^{\nu}}{ k^2 -\xi_{i} 
M_{i}^2}\Bigg ]D_{i}(k)
\label{GenProp}
\ee
with
\be
D_{i}(k)= (k^2 -M_{i}^2)^{-1}
\label{Deno}
\ee
where $i=W,V$  with $V =  Z,A$ and $M^2_{A}=0$;  $k$ denotes
the virtual four-momentum circulating in the loop.  Clearly,  in the case of
$\Delta^{\mu\nu}_{i} (k)$  the  longitudinal momenta are   those
proportional to $(1-\xi_{i})$.
The longitudinal terms arising from the tri-linear vertex 
may be identified by splitting its Lorentz structure 
$\Gamma_{\alpha\mu\nu}(q,-k,k-q)$
appearing inside the one-loop diagrams (where $q$ denotes 
the physical four-momentum entering into the vertex, 
see Section~\ref{expl}) into two parts~\cite{Cornwall:1976ii}:
\bea
\Gamma_{\alpha\mu\nu} (q,-k,k-q) 
&=& (k+q)_{\nu} g_{\alpha\mu} + (q-2k)_{\alpha} g_{\mu\nu} 
(k-2q)_{\mu} g_{\alpha\nu} \nonumber\\  
&=& \Gamma_{\alpha\mu\nu}^{\rm F} + \Gamma_{\alpha\mu\nu}^{\rm P} \, ,
\label{decomp}
\eea
where
\bea
\Gamma_{\alpha\mu\nu}^{\rm F}&=& 
(q-2k)_{\alpha} g_{\mu\nu} + 2q_{\nu}g_{\alpha\mu} 
- 2q_{\mu}g_{\alpha\nu}, \nonumber\\
\Gamma_{\alpha\mu\nu}^{\rm P} &=&
 (k-q)_{\nu} g_{\alpha\mu} +k_{\mu}g_{\alpha\nu}.  
\label{GFGP}
\eea
The first term in $\Gamma_{\alpha\mu\nu}^{\rm F}$ is a convective
vertex describing the coupling of a vector boson to a scalar field, 
whereas the other two terms originate from spin or magnetic moment. 
The above decomposition assigns a special r\^ole 
to the $q$-leg,
and allows $\Gamma_{\alpha\mu\nu}^{\rm F}$ 
to satisfy the Ward identity
\be 
q^{\alpha} \Gamma_{\alpha\mu\nu}^{\rm F}= 
 (k-q)^2 g_{\mu\nu} - k^2  g_{\mu\nu}\, ,
\label{WI2B}
\ee
The relevant Ward identities triggered
by the longitudinal momenta identified above, are then two. The first one reads
\bea
\lslush  \PrL &=& (\lslush + \pslush ) \PrL 
- \PrR \not\! p 
\nonumber\\
&=& S_{f'}^{-1}(\lslush + \pslush ) \PrL - 
P_{ R} S_{f}^{-1}(\pslush) + m_{f'} \PrL  -  m_{f} \PrR,
\label{EWI1}
\eea
where $P_{R(L)} = [1  + (-) \gamma_5]/2$  is the  chirality projection
operator and   
$iS_{f}$ is the tree-level propagator of the fermion $f$;
$f'$ is the
iso-doublet partner of the external fermion $f$.
(Alternatively, one may adopt
the formulation of the PT in terms of equal-time commutators 
of currents \cite{Degrassi:1992ue}).
The second Ward identity reads 
\be
(k-q)^{\nu}\Gamma_{\alpha\mu\nu}(q,-k,k-q) = 
\bigg[k^2 g_{\alpha\mu} - k_{\alpha}k_{\mu}\bigg]-
\bigg[q^2 g_{\alpha\mu} - q_{\alpha}q_{\mu}\bigg],
\label{EWI2}
\ee
together with the Bose-symmetric one, when contracting with $k^{\mu}$ instead of
$(k-q)^{\nu}$.

The appearance of inverse propagators leads (through the pinching out 
of the corresponding internal propagators) to 
$\xi$-dependent contributions which are topologically distinct from those of their 
parent Feynman graph.
Because of that, 
all $\xi$-dependent parts violate one or more of the 
properties ({\it i})--({\it x}) listed in the Introduction.  
In particular, if $m_i=0$
the gauge-dependent parts are purely propagator-like, {\it i.e.}, they violate 
({\it vi}), whereas if  $m_i \neq 0$ they are either propagator-like 
(hence violating  ({\it vi})), or they are multiplied by a factor $q^2$, {\it i.e.}, they effectively violate~({\it i}). 

\begin{figure}[t]
\includegraphics[width=12cm]{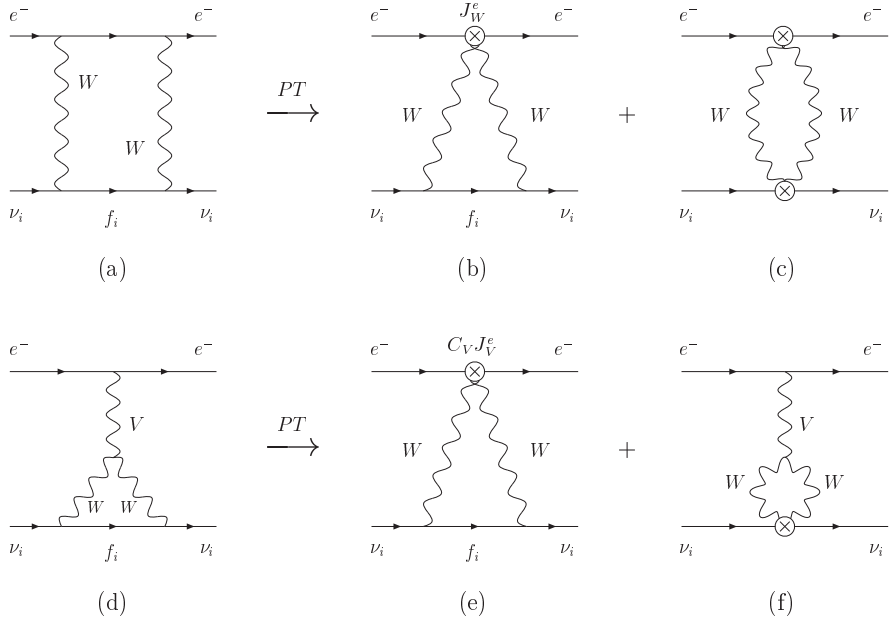}
\caption{The general topology of the relevant gauge-dependent contributions in the presence 
of non-vanishing fermion masses. The contributions (b) and (e) are proportional to $m_i$.
If the target-fermions are (massless) right-handed fermions, 
the entire first column of the right-hand side vanishes.}
\end{figure}

Before entering into the explicit proof, it is instructive to   
briefly sketch how the cancellation of the gauge-dependent terms proceeds:
\begin{itemize}
\item[$\bullet$] The $\xi$-dependent  propagator-like (universal, or equivalently 
flavor-independent) parts [shown in (c) and (f)] 
cancel against corresponding $\xi$-dependent contributions from the 
conventional self-energy graphs (not shown), for every neutrino flavor. 
The emerging gauge-independent effective self-energies form renormalization-group-invariant 
quantities (electroweak effective charges) exactly as in QED.
\item[$\bullet$] The vertex-like (flavor-dependent) 
parts [shown in (b) and (e)] are proportional
to $m_i$; they stem from the mass-terms appearing on the right-hand side of Eq.(\ref{EWI1}). 
The particular topology of the graph (e)
({\it i.e.}, contact-type interaction) 
arises because the  
$\xi$-dependent part coming from the $A^{*}\nu \nu$ vertex 
are
proportional to $q^2$, whereas those from the  $Z^{*}\nu \nu$ vertex 
are proportional to $(q^2-M_Z^2)$ [this is essentially due to the Ward identity of Eq.(\ref{EWI2})].
On the other hand, graph (b) has this form due to Eq.(\ref{EWI1}).
The topological form of these graphs is crucial 
for the cancellation, because it allows parts of the three (originally distinct) graphs
to ``talk'' to each other.
The next step is to recognize that, 
the current structures of the contact-like interactions are such 
that the total sum of the two pieces is zero, {\it i.e.}, 
\be
({\rm b})+\sum_V({\rm e})=0.
\ee
This final cancellation is not accidental, but a direct 
consequence of the current operator 
identity of Eq.(\ref{fci}).
\end{itemize}

We emphasize that (see also property ({\it vi}) in the Introduction) 
the definition and value of NCR should be independent 
of the charged probe used. 
If one were to use as a probe massless right-handed electrons, $e_{R}$,   
there would be no boxes containing $W$-bosons involved in the 
gauge cancellations, $({\rm b})=0$; however in this case the current 
structures of diagrams $({\rm e})$  get modified in such a way that now 
$\sum_V({\rm e})=0$ (see \cite{Bernabeu:2000hf} and below for a detailed discussion).

\subsection{Review of the massless case}

The massless case has been studied in 
detail in \cite{Bernabeu:2000hf,Bernabeu:2002nw}. 
For  concreteness, we  will focus  on  the same  process 
considered there, namely
the (one-loop) elastic scattering
process    $e(\ell_1)\nu(p_1)\to e(\ell_2)\nu(p_2)$,    with the
Mandelstam  variables   defined  as  $s=(\ell_1+p_1)^2=(\ell_2+p_2)^2$,
$t=q^2=(p_1-p_2)^2=(\ell_2-\ell_1)^2$,    $u=(\ell_1-p_2)^2=(\ell_2-p_1)^2$,   and
$s+t+u=0$. Notice that $\nu$ will be chosen
to belong in a different   
iso-doublet than the target-electron 
(the muon neutrino $\nu_\mu$, for example),  so that  the crossed  (charged)
channel vanishes. 
According to the general PT algorithm, the one-loop amplitude 
for the above process may be  reorganized into sub-amplitudes 
which have the same kinematic properties as self-energies, vertices, and boxes,
and are at the same time completely independent of the gauge-fixing parameter.
One of these sub-amplitudes will be identified with the 
one-loop 
effective photon-neutrino vertex.
 
What has been assumed in \cite{Bernabeu:2000hf,Bernabeu:2002nw,Bernabeu:2002pd}  when constructing 
the aforementioned vertex  
is that the masses of the fermions 
may be neglected when carrying out the various cancellations;
in particular, one employs the Ward identity 
of Eq.(\ref{EWI1}), with the masses set to zero.
In such a case, as mentioned earlier, all contributions 
stemming from longitudinal momenta are effectively propagator-like;
they are combined with the normal self-energy graphs giving rise to 
two renormalization group invariant quantities, corresponding to the 
electroweak  effective charge  and the  running  mixing angle 
\cite{Hagiwara:1994pw}.

After having removed all gauge-dependent propagator-like 
contributions, the remaining genuine one-loop  vertex, 
to be denoted by $\widehat{\Gamma}_{A_\mu \nu_i \nu_i}$,
is completely independent of the gauge-fixing parameter, and 
in addition satisfies a naive, QED-like Ward identity. 
As explained in detail in the literature,
the final answer is given by the two 
graphs of Fig.2, where the Feynman gauge 
is used for all internal gauge-boson propagators,
and the usual tree-level three-boson vertex $\Gamma_{\alpha\mu\nu}$
is replaced by $\Gamma_{\alpha\mu\nu}^{\rm F}$.

\begin{figure}[t]
\includegraphics[width=10cm]{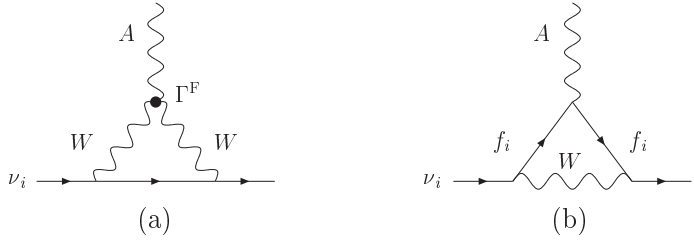}
\caption{\label{NCR}The diagrams contributing to the PT neutrino charge radius.
Here $\Gamma^{\rm F}$~stands for the Feynman part of the tri-linear gauge-boson
vertex, which is defined in Eq(\ref{decomp}). Notice that the logarithmic term in 
the NCR expression of Eq.(\ref{ncr}) originates entirely from the Abelian-like diagram (b).}
\end{figure}

It is straightforward to evaluate
the two aforementioned vertex graphs; their  
sum gives a ultra-violet finite result, from which 
one can extract the 
dimension-less
electromagnetic form-factor $\widehat{F}_1^{A}(q^2)$.  
In particular, since $\widehat{F}_1^{A}(q^2)$ is 
proportional to $q^2$, we may define the dimension-full
form-factor $\widehat{F}_{\nu_i}(q^2)$ as
\be
\widehat{\Gamma}_{A_\mu \nu_i \bar{\nu}_i}
= \widehat{F}_1^{A}(q^2) \left[ie \gamma_{\mu}(1-\gamma_{5})\right]
= q^2  \widehat{F}_{\nu_i}(q^2)
\left[ie \gamma_{\mu}(1-\gamma_{5})\right],
\label{TheFS}
\ee
from which the NCR is defined in the usual way as $\widehat{F}_{\nu_i}(0)=\frac16\langle 
r^2_{\nu_i}\rangle$.
We emphasize that, when taking the limit 
$q^2\to 0$, as dictated by the very definition of the NCR,  
$\widehat{F}_{\nu_i}(q^2)$ is infrared divergent, unless 
the mass $m_i$ of the
charged iso-doublet
partner of the neutrino is kept non-zero. 
In particular, when  $m_i \to 0$ 
a logarithmically divergent contribution emerges from the 
Abelian-like diagram of Fig.\ref{NCR}(b). 
Thus, a non-zero mass must be eventually kept in the 
calculation of the final answer even in the ``massless'' case; 
evidently, the term 
``massless'' refers to the fact that the massless version 
of the Ward identity has been employed, and mass terms appearing in the
numerators (but not the denominators)
of the corresponding Feynman graphs have been discarded.  

The final answer is given by~\cite{Bernabeu:2000hf}
\be
\big <r^2_{\nu_i}\,  \big> =
\frac{G_{\rm F}}{4\, {\sqrt 2 } \pi^2} 
\left[3 
- 2\log \left(\frac{m_{i}^2}{M_{W}^2} \right) \right],
\qquad i= e,\mu,\tau
\label{ncr}
\ee
where $G_{\rm F} = \gw^2 \sqrt{2}/8 M_{W}^2$ is the 
Fermi constant and $\gw=e\sw$ the $SU(2)$ gauge coupling.
The numerical values obtained for the corresponding NCR of the three 
neutrino families \cite{Bernabeu:2002nw,Bernabeu:2002pd}
are consistent with various bounds that have appeared 
in the literature \cite{Salati:1994tf}. 

\subsection{\label{expl}The massive case: explicit one-loop calculation}

After these introductory remarks, in  the rest of this section we will
show explicitly how the crucial cancellations, enforced by the
PT through the  tree-level Ward identities of the  theory, continue to
hold even when we relax the hypothesis of working with purely massless
fermions.
We will  prove this in two  different ways: first we  will carry out an
explicit one-loop calculation following simply the PT rules; second,
we  will  re-do the  analysis  of  gauge  cancellations in  their  full
generality, through  the use of  the so-called NIs, and
show that the latter lead precisely to the same kind of rearrangements
induced by the PT.

We consider again the same reference process $e(\ell_1)\nu(p_1)\to e(\ell_2)\nu(p_2)$ 
as before, and carry out the PT-rearrangement of the corresponding 
one-loop amplitude, but this time we will maintain throughout the 
calculation non-vanishing masses for the iso-doublet partner of the 
neutrino. 
In particular, we  are interested in analyzing the 
box/vertex gauge cancellation in the case where
we do not neglect the masses of the fermions propagating inside
the loop,  as originally done  in \cite{Bernabeu:2002nw}. 
Notice that the PT rearrangement, in addition to the 
quantities relevant for defining the NCR, will also give rise 
to one-loop vertices involving the off-shell photon or $Z$-boson 
and the target-fermions. In order to simplify the picture
we will assume the target fermions to be massless. This has no bearing
whatsoever on the cancellations taking place in the loops 
containing the  massive iso-doublet partner of the neutrino; in any case,  
the assumption of massless target fermions 
will be relaxed later on, when presenting the treatment based on 
the NIs.
Therefore, in what
follows we will only focus on the PT terms contributing to the gauge
cancellation  of the  photon-neutrino  vertex $\Gamma^{(1)}_{A_\mu\nu\nu}$,
and will not display terms involved in other parts of the full 
PT rearrangement, as for example in the construction of the 
one-loop vertex involving the target fermions.
In addition, from the terms  
involved in  the construction of the  photon-neutrino  vertex 
we will display only those 
proportional
to the mass of the iso-doublet partner, since these are the new terms 
not considered in  \cite{Bernabeu:2002nw}.

Let us start by introducing the relevant tree-level photon and $Z$-boson vertices as
\be
\Gamma^{(0)}_{V_\mu \bar ff}=i\gw\gamma_\mu C_V^f,
\qquad\quad
C_V^f=\left\{
\begin{array}{ll}
-\sw Q_f, &\quad {\rm if}\ \ V=A, \\
-\frac1\cw
\left[\sw^2 Q_f-T^{f}_{z}P_L\right], & \quad {\rm if}\ \ V=Z.
\end{array}
\right.
\ee
where $Q_f$ is the electric charge of the fermion $f$, $T^f_{z}$ represents 
the $z$ component of the weak iso-spin [which is $+(-) 1/2$ for up (down)-type leptons],
and $\cw=\sqrt{1-\sw^2}=M_W/M_Z$. 
In addition we define the following integrals
\bea
I_1&=&\{(k^2-\MSW)(k^2-\xiw \MSW)[(k-q)^2-\MSW]\}^{-1},  \nonumber \\
I'_1&=&\{(k^2-\MSW)(k^2-\xiw \MSW)[(k-q)^2-\xiw \MSW]\}^{-1},  \nonumber \\ 
I_2&=&\{(k^2-\MSW)[(k-q)^2-\MSW] [(k-q)^2-\xiw \MSW]\}^{-1},  \nonumber \\
I'_2&=&\{(k^2-\xiw \MSW)[(k-q)^2-\MSW] [(k-q)^2-\xiw \MSW]\}^{-1},  \nonumber \\     
I_3&=&\{(k^2-\MSW)(k^2-\xi_{  W}\MSW)[(k-q)^2-\MSW][(k-q)^2-\xiw \MSW]\}^{-1},  \nonumber \\
I_4&=&\{(k^2-\MSW)[(k-q)^2-\xiw \MSW]\}^{-1},  \nonumber \\ 
I_5&=&\{(k^2-\xiw \MSW)[(k-q)^2-\MSW]\}^{-1},\nonumber \\ 
I_6&=&\{(k^2-\MSW)(k^2-\xiw \MSW)\}^{-1},
\eea
together with the characteristic PT structures
\bea
J&=&\bar u_{\nu}(p_2)\PrR\left[m_{\mu}^2S_\mu(\lslush+\pslush_1)\right]
\PrL u_{\nu}(p_1), \nonumber \\
J^\alpha_1&=&\bar u_{\nu}(p_2)\gamma^\alpha \PrL \left[m_{\mu}S_\mu(\lslush+\pslush_1)\right]
\PrL u_{\nu}(p_1), \nonumber \\
J^\alpha_2&=&\bar u_{\nu}(p_2)\PrR\left[m_{\mu}S_\mu(\lslush+\pslush_1)\right]\gamma^\alpha
\PrL u_{\nu}(p_1).
\eea
\begin{figure}[t]
\includegraphics[width=15cm]{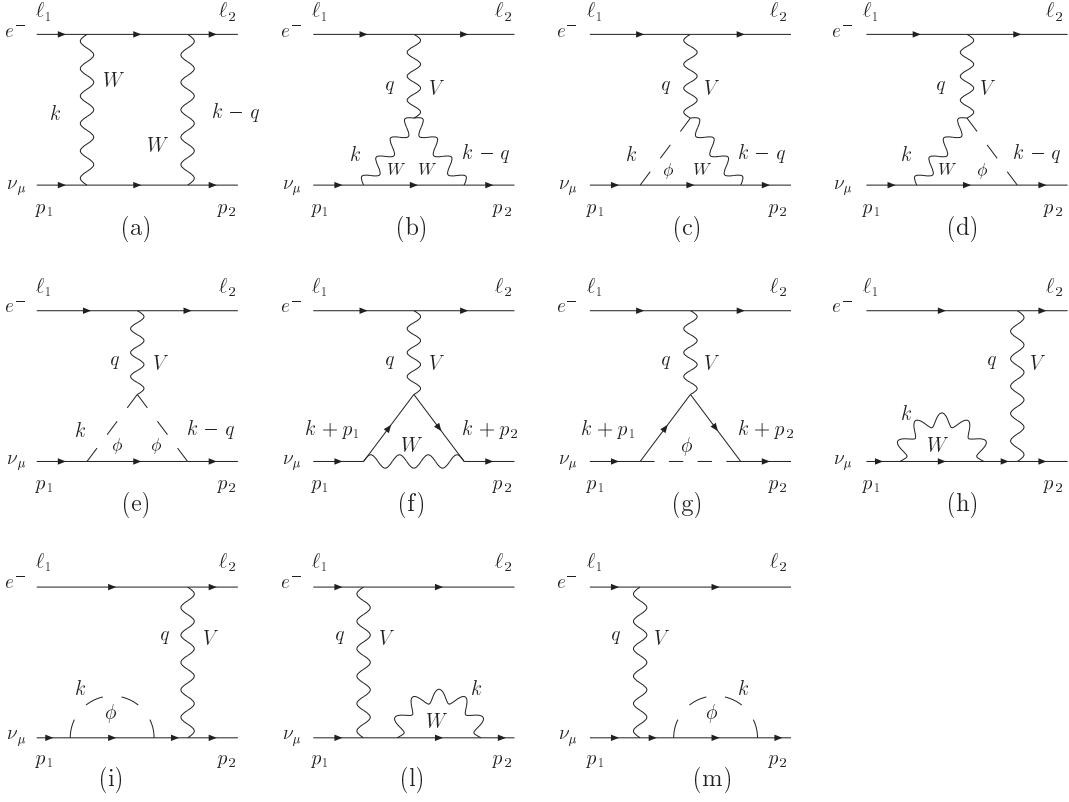}
\caption{\label{alldiagrams} The diagrams contributing to the box/vertex PT 
cancellation for the case of massive propagating fermions.}
\end{figure}
Turning to individual diagrams, we first consider the box diagram
$(a)$ of Fig.\ref{alldiagrams}, and isolate 
the PT vertex-like contributions;
they will eventually cancel against the corresponding gauge dependent 
contributions of the vertices
$\Gamma^{(1)}_{V_\mu\nu\nu}$. Suppressing the factor
$\frac{(i\gw)^3}2\int\!\frac{d^dk}{(2\pi)^n}$, with  
$d=4-\epsilon$ the space-time dimension, one has
\be 
({\rm a})= J^{e}_{W_\alpha} \left[-\lw I_1 J^\alpha_1
-\lw I_2 J^\alpha_2
+\lw^2 I_3 (k-q)^\alpha J\right] 
\label{BV}
\ee
where $\lw=1-\xiw$, and the current $J^{e}_{W_\alpha}$ is defined according to
\be
J^e_{W_\alpha}=-i\frac\gw2\bar
u_{e}(\ell_2)\gamma_\alpha \PrL 
u_{e}(\ell_1),
\ee
and should not be confused with
the usual $W$ current connecting up- and down-type fermions. 

For the vertex graph (b) there are quite a few terms to take into
account with respect to the massless case; 
one has, in fact
\bea
({\rm b})&=&-\lw I_1C_V\JV D_V(q)\left\{(q^2-M_V^2)+M^2_V-[(k-q)^2-M^2_W]
-M^2_W\right\}J^\alpha_1\nonumber \\
&-&\lw I_2C_V\JV D_V(q)\left\{(q^2-M_V^2)+M^2_V-[(k-q)^2-M^2_W]
-M^2_W\right\}J^\alpha_2
\nonumber \\
&-&\lw C_V\JV D_V(q)\left\{I_1(k-q)^\alpha+I_2k^\alpha +
i\lw I_3[(q^2-M_V^2)+M^2_V]k^\alpha\right\}
(k-q)^\alpha J,\nonumber \\
\label{(b)}
\eea
where $D_V(q)$ is as in Eq.(\ref{Deno}), 
$C_V=\sw$ (respectively, $C_V=-\cw$) when $V=A$ (respectively $V=Z$), and
the current $\JV$ is defined according to
\be
\JV=i\gw\bar
u_{e}(\ell_2)\gamma_\alpha C_V^e 
u_{e}(\ell_1).
\ee
The twelve terms appearing in Eq.(\ref{(b)}) will be referred to as $({\rm b}_1)$, $({\rm b}_2)$ and so on (the same kind of numbering will be adopted for all the following expressions possessing more than one term).
Next let us consider diagrams (c) and (d); the longitudinal components of the internal $W$  
can trigger elementary Ward-identities, and give rise to pinch contributions.
(the vertex diagram (e) cannot possibly pinch).
Thus, from  (c) and (d) 
we find
\bea
({\rm c})&=&I_4C'_V\JV D_V(q)J^\alpha_2-\lw I'_1C'_V\JV D_V(q)k^\alpha J,\nonumber \\
({\rm d})&=&I_5C'_V\JV D_V(q)J^\alpha_1-\lw I'_2C'_V\JV D_V(q)(k-q)^\alpha J,
\eea
where $C'_V=\sw$ (respectively, $C'_V=\sw^2/\cw$) when $V=A$ (respectively, $V=Z$). The gauge dependent pieces above, combine then with some of the ones appearing in Eq.(\ref{(b)}), to give rise to the following gauge-independent combinations
\bea
&&({\rm c}_1)+({\rm b}_4)+({\rm b}_8)=\left.({\rm c})\right|_{\xiw=1},\nonumber \\
&&({\rm d}_1)+({\rm b}_2)+({\rm b}_6)=\left.({\rm d})\right|_{\xiw=1},\nonumber \\
&&({\rm e})+({\rm c}_2)+({\rm d}_2)+({\rm b}_9)+({\rm b}_{10})+({\rm b}_{12})=\left.({\rm e})\right|_{\xiw=1}.
\eea

Finally, we have to consider the diagrams (f) through (m). For the abelian-like vertex (f) we find
\bea
({\rm f})&=&-\lw I_6\JV\bar u_\nu\PrR\left\{\gamma^\alpha C^e_V m_{\mu}S_\mu(\lslush+\pslush_1)
+m_{\mu}S(\lslush+\pslush_2)\gamma^\alpha C^e_V\right.\nonumber \\
&+&\left. m_{\mu}S_\mu(\lslush+\pslush_2)\gamma^\alpha C^e_V m_{\mu}S_\mu(\lslush+\pslush_1)
\right\}\PrL u_\nu,
\eea
while for the wavefunction renormalization graphs we obtain
\bea
({\rm h})&=&\lw I_6\JV D_V(q)\bar u_\nu\gamma^\alpha C^\nu_V\PrL m_{\mu}S_\mu(\lslush+\pslush_1)\PrL u_\nu\nonumber \\
&-&\lw I_6\JV D_V(q)\bar u_\nu\gamma^\alpha C_V^\nu S_\nu(\pslush_2)\PrR
m_{\mu}^2S_\mu(\lslush+\pslush_1)\PrL u_\nu, \nonumber \\
({\rm l})&=&\lw I_6\JV D_V(q)\bar u_\nu\PrR m_{\mu}S_\mu(\lslush+\pslush_2)\PrR
\gamma^\alpha C^\nu_V u_\nu \nonumber \\
&-&\lw I_6\JV D_V(q)\bar u_\nu\PrR m_{\mu}^2
S_\mu(\lslush+\pslush_2)\PrL S_\nu(\pslush_2)\gamma^\alpha C_V^\nu u_\nu.
\eea
All remaining diagrams are inert as far as the PT rearrangement is concerned.
Putting them all together one obtains 
the following gauge-independent combinations
\bea
&&({\rm f}_3)+({\rm g})=\left.({\rm q})\right|_{\xiw=1},\nonumber \\
&&({\rm h}_2)+({\rm i})=\left.({\rm i})\right|_{\xiw=1},\nonumber \\
&&({\rm l}_2)+({\rm m})=\left.({\rm m})\right|_{\xiw=1},
\eea
together with the cancellation
\be
({\rm f}_1)+({\rm h}_1)+({\rm b}_3)=({\rm f}_2)+({\rm l}_1)+({\rm b}_7)=0.
\ee

Therefore the gauge-dependent terms left-over after summing over all 
conventional vertex graphs [(b) through (m)] reads 
\be
({\rm b})_{\rm{lo}}=\sum_V C_V \JV \left\{-\lw I_1  J^\alpha_1
-\lw I_2 J^\alpha_2
+\lw^2 I_3 (k-q)^\alpha J\right\}.
\label{DB}
\ee
Note that this term is already of the contact type, 
{\it i.e.}, the $1/q^2$ or $1/(q^2-M_Z^2)$ tree-level
propagator have been cancelled out,    
before carrying out the 
integration over the virtual momentum $k$. 
The crucial point, central in the PT philosophy, 
 is that the same vertex-like contact term 
has already been identified being concealled inside the box, {\it viz}. 
Eq.(\ref{BV}), and should therefore form part of the definition of the 
effective photon-neutrino vertex. Indeed, after employing the 
current relation 
\be
J^e_{W_\alpha}=-\sum_V C_V\JV,
\label{CI}
\ee
it is clear that the aforementioned contribution couples electromagnetically to 
the target; thus the two gauge-dependent pieces should be naturally added, 
yielding
\be
({\rm a})+({\rm b})_{\rm{lo}}=0.
\ee
Of course, Eq.(\ref{CI}) is nothing but 
the current relation of Eq.(\ref{fci}), modulo the following current redefinitions
\be
2J^e_{W_\alpha}=i\gw J^{(3)}_\alpha, \qquad
J^e_{A_\alpha}=-i\gw\sw J_{\gamma_\alpha}, \qquad
\cw J^e_{Z_\alpha}=i\gw J_{Z_\alpha}.
\ee

It is important to realize that, had we chosen the 
target electrons to be massless and 
right-handedly polarized (as originally done in \cite{Bernabeu:2000hf}), 
then $({\rm a})=0$ right from the beginning (there are no $WW$ boxes in such case), 
but the cancellation would proceed in the very same way, since the current relation 
above is modified to read
\be
\sum_V C_VJ^{e_{R}}_{V_\alpha}=0.
\ee
Thus, it becomes clear that the NCR defined through this procedure 
is identified with a quantity {\it independent}   
of the particle or source used to probe it.
What depends on the details 
of the target
is only the precise way that the various  
diagrammatic contributions 
conspire in order to always furnish the same unique and gauge-independent 
answer (for example the presence or absence of $WW$ boxes).

As expected, the simplifying hypothesis of neglecting 
the mass of the (propagating) electron does not conflict with the PT algorithm.
Evidently, 
the fate of the gauge-dependent terms 
proportional to $(m_i^2/M_W^2)$ stemming from the 
conventionally defined $A^{*}\nu \nu$ vertex [see Eq.(\ref{DB})], 
is exactly the same as that of their massless counter-parts:
they completely cancel against exactly analogous terms stemming from 
the box, 
before any integration over the virtual momentum is carried out.
Therefore, any potentially pathological 
behavior of those terms for special values of $\xi$ is absolutely immaterial.
In particular, 
the straightforward (but really unnecessary)
integration of the terms given in Eq.(\ref{DB}) would
yield a contribution 
$\frac{G_{\rm F}}{192\, {\sqrt 2 } \pi^2} \frac{m_i^2}{M_W^2} 
\frac{1}{\lambda_W}\left[10 (1-\xiw)+(9+\xiw)\log\xiw\right]$
to the conventionally defined NCR, which diverges  as $\xi\to0$ \cite{Fujikawa:2003ww},
or $\xi\to\infty$.
In conclusion, after the PT cancellation procedure has been 
completed, any gauge-dependence  
disappears, and all contributions proportional to $m_i^2/M_W^2$ 
are therefore
genuinely suppressed (numerically), and cannot be made arbitrarily large.

\begin{figure}[t]
\hspace{-0.6cm}
\includegraphics[width=16.0cm]{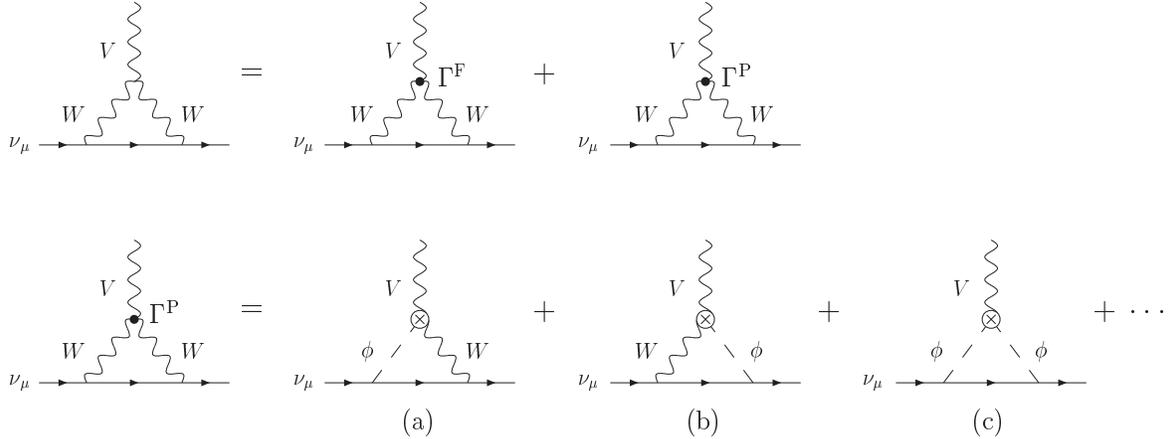}
\caption{\label{PTgbsector}Carrying out the PT procedure for constructing the neutrino charge radius.
As a result of the Slavnov-Taylor identities triggered by the longitudinal 
pinching momenta contained in $\Gamma^{\rm P}$, one obtains the PT terms (a), (b) and (c) that, 
when added to the corresponding $R_\xi$ ones, will project the theory to the Feynman gauge of the background field method. 
The dots denote propagator like terms.}
\end{figure}

The last step in the 
the  construction of the one-loop PT $\widehat{A}^{*}\nu \nu$ vertex
is to  carry  out  the  characteristic   PT  decomposition  of
Eq.(\ref{EWI2})  on  the  triple gauge  boson  vertex  $AWW$
(Fig.\ref{PTgbsector}), which contains the last remaining 
longitudinal momenta.
All  other diagrams  will
be inert as far as this final 
PT rearrangement is concerned, simply because 
they do not possess  longitudinal pinching momenta any more.
The  longitudinal momenta
contained  in  the  $\Gamma^{\rm  P}$  term, will  next  trigger  some
suitable Slavnov-Taylor identities \cite{Binosi:2002ft,Binosi:2004qe}.
This triggering will produce  additional vertex-like PT pieces, which,
finally, combine with the inert diagrams, and conspire to rearrange the
perturbative  series  in  a  very  precise way: in  our
particular   case,  the   aforementioned  PT   terms  will   contain  a
contribution  that cancels  exactly  the one  coming  from the  vertex
$AW\phi$ [Fig.\ref{PTgbsector}(a)  and (b)], and one  that will modify
the         triple         gauge-boson--scalar--scalar        coupling
[Fig.\ref{PTgbsector}(c)].  The final outcome of this procedure is the
rearrangement  of the  perturbative  series in  such  a way  as to  be
dynamically  projected to the  Feynman gauge  of the  background field
method; this latter  fact is not limited to  the present one-loop NCR,
but  it has  been proved  to be  valid to  all orders  in the  full SM
\cite{Binosi:2004qe}.

\section{\label{aoc}Nielsen identities analysis}

Recently \cite{Binosi:2002ft,Binosi:2004qe}, it has been realized that 
the fundamental underlying symmetry which is driving 
the PT cancellations is the 
Becchi-Rouet-Stora-Tyutin (BRST) symmetry \cite{Becchi:1976nq}. This  
realization has been instrumental in generalizing the PT procedure 
to all orders, and has allowed for its connection to 
powerful BRST related formalisms. 
As a result, one can take full advantage 
of the BRST symmetry to deeper analyze 
the nature of the PT cancellations discussed in our one-loop example, in general, 
and the issue of neglecting terms proportional to the 
masses of the internal fermions, in particular. 
To this end, we will work within the framework of the
Batalin-Vilkovisky~\cite{Batalin:1977pb} and Nielsen~\cite{Nielsen:1975fs} formalisms,
which will be briefly reviewed in what follows.

\subsection{General formalism}

\begin{figure}[t]
\includegraphics[width=14cm]{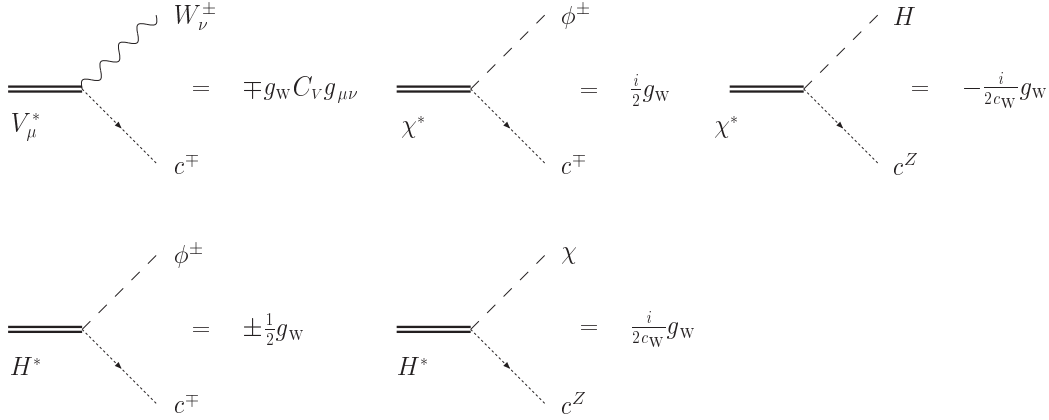}
\caption{\label{BRSTfr} The Feynman rules for the neutral 
gauge bosons ($V^*_\mu$) and scalars ($\chi^*$, $H^*$) BRST sources coming from the Lagrangian 
${\cal L}_{\rm BRST}$.}
\end{figure}

As far as the Batalin-Vilkovisky formalism is concerned, 
one introduces for each SM field $\Phi$, the corresponding anti-field $\Phi^*$, and couples them through the
Lagrangian (for details see also \cite{Grassi:1999tp,Binosi:2002ez})
\be
{\cal L}_{\rm BRST}=\sum_\Phi\Phi^* s\,\Phi,
\label{LBRST}
\ee
where $s$ is the BRST operator. In particular, the coupling of the neutral 
gauge bosons and scalar BRST sources (denoted by 
$V^*_\mu$, and $\chi^*$, $H^*$ respectively), read
\bea
{\cal L}_{\rm BRST}&\supset& i\gw C_V V^*_\mu(W^{+\mu}c^--W^{-\mu}c^+)
+\frac\gw2\chi^*(\phi^+c^-+\phi^-c^+)-\frac\gw{2\cw}\chi^*Hc^Z\nonumber \\
&+&\frac{i\gw}2H^*(\phi^+c^--\phi^-c^+)+\frac\gw{2\cw}H^*\chi c^Z,
\eea
$c^\pm$ and $c^V$ being the charged and neutral ghost fields, respectively.
Then from ${\cal L}_{\rm BRST}$, one finds the Feynman rules shown in Fig.\ref{BRSTfr}.

The BRST invariance of the SM action, or, equivalently, the
unitarity of the $S$-matrix and the gauge independence of the physical
observables, are then encoded into the master equation
\be
{\mathfrak S}(\Gamma)=0,
\label{me0}
\ee
where
\be
{\mathfrak S}(\Gamma)=\int\!d^4x\,\sum_\Phi\frac{\delta^R\Gamma}{\delta\Phi}
\frac{\delta^L\Gamma}{\delta\Phi^*}.
\label{me}
\ee
In Eq.(\ref{me}), the sum runs over all the SM fields, $R$ and $L$
denote the right and left differentiation, respectively, and 
$\Gamma$ represents the effective action [which depends on the antifields
through Eq.(\ref{LBRST})]. 
This equation can be used to derive the complete set of
non-linear STIs to  all orders in the perturbative theory, via the
repeated application of functional differentiation. 

However the important point here is that by enlarging the BRST symmetry of the theory,
one can construct a tool that allows  
to control the dependence of the Green's functions on the gauge parameter $\xi_i$ (with $i={W,V}$)
in a completely algebraic way.
In fact, let us promote the parameters $\xi_i$ to (static) fields, and introduce their 
corresponding BRST sources $\eta_i$, in such a way that
\be
s\xi_i=\eta_i,\qquad\qquad s\eta_i=0.
\label{ext}
\ee

After doing this, the BRST invariance of the ghost (${\cal L}_{{\rm FPG}}$) and gauge 
fixing (${\cal L}_{{\rm GF}}$) sectors of the SM Lagrangian is lost, and to 
restore it one has to add to
the the sum ${\cal L}_{{\rm GF}}+{\cal L}_{{\rm FPG}}$ the term
\be
{\cal L}_{\rm N}= -\frac1{2\xiw}\etaw\left(\bar c^+{\cal F}^++\bar
c^-{\cal F}^-\right)-\sumv\frac1{2\xi_V}\eta_V\left(\bar c^V
{\cal F}^V\right),
\ee
where ${\cal F}^\pm$, ${\cal F}^V$ represent the gauge fixing functions, {\it i.e.},
\be
{\cal F}^\pm
=\partial^\mu W^\pm_\mu\mp i\xi_{ W}\Mw\phi^\pm,
\qquad{\cal F}^{V}=
\partial^\mu V_\mu- \xi_{V}M_V\chi,
\ee
within the class of $R_\xi$ gauges used in our analysis.
\begin{figure}[!t]
\includegraphics[width=10.5cm]{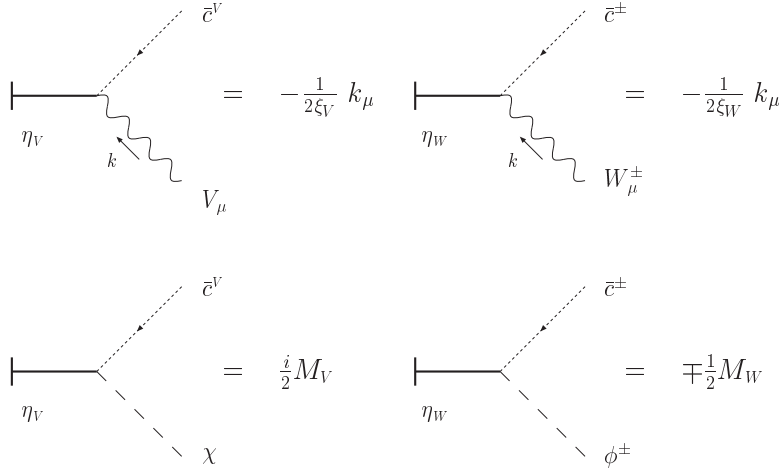}
\caption{\label{NIfr} The Feynman rules for the $\eta_i$ static sources  
coming from the Lagrangian ${\cal L}_{\rm N}$.}
\end{figure}
The term ${\cal L}_{\rm N}$ will then control the couplings of the sources $\eta_i$ to 
the SM fields, giving rise to the Feynman rules shown in Fig.\ref{NIfr}.
For all practical calculations one can set $\eta_i=0$ thus recovering
both 
the unextended BRST transformations,
as well as the master equation of Eq.(\ref{me0}). However when $\eta_i\neq0$, the master equation reads
\be
{\mathfrak S}_{\eta_i}(\Gamma)=0,
\ee
where
\be 
{\mathfrak S}_{\eta_i}(\Gamma) = {\mathfrak
S}(\Gamma)+\eta_i\partial_{\xi_i}\Gamma.
\ee
Thus, after differentiating 
this new master equation
with respect to $\xi_i$, and setting $\eta_i$
to zero, we obtain
\be
\left.\partial_{\xi_i}\Gamma\right\vert_{\eta_i=0}=-\left.\left(
\int\!d^4x\,\partial_{\eta_i}\sum_\Phi\frac{\delta^R\Gamma}{\delta\Phi}
\frac{\delta^L\Gamma}{\delta\Phi^*}\right)\right\vert_{\eta_i=0}.
\label{NI}
\ee
Establishing the above functional equation, allows, via the
repeated application of functional differentiation, to derive a set of identities,
known in the literature under the name of NIs~\cite{Nielsen:1975fs},
that control the
gauge-parameter(s) dependence of the different Green's functions
appearing in the theory. Therefore, NIs can be used to unveil 
in their full generality 
the patterns of gauge cancellations
occurring inside gauge independent 
quantities such as $S$-matrix elements. In fact, 
as has been demonstrated recently \cite{Binosi:2004qe},
these patterns are actually of the PT type.
Notice, however, that, unlike the PT, NIs 
cannot be used to construct gauge invariant and 
gauge-fixing-parameter-independent Green's functions. 

Finally, a technical remark. The extension of the BRST symmetry through
Eq.(\ref{ext}) is just a technical trick to gain control over the 
gauge-parameter dependence of the various Green's functions appearing in the
theory; thus, unlike the STIs generated from Eq.(\ref{me}),
Eq.(\ref{NI}) does not have to be preserved in the renormalization
procedure, which will in general deform it (see \cite{Gambino:1999ai} and
references therein). 
The complications due to this fact may be circumvented
by choosing to work within a renormalization scheme 
that fixes the parameters of ${\cal L}_{\rm SM}^{\rm cl}$ using physical observables 
(see again~\cite{Gambino:1999ai}).
\begin{figure}[t]
\includegraphics[width=13cm]{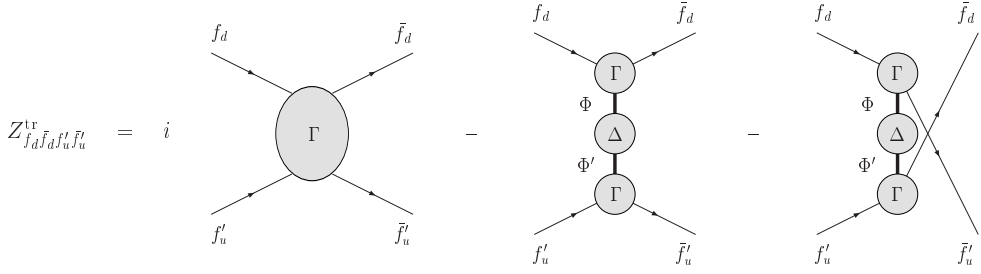}
\caption{\label{Zdeco} The decomposition of the
truncated Green's function $Z_{f_d\bar f_df'_u\bar f'_u}^{\rm tr}$. Notice that the last
diagram is not present when $f_d$ and $f'_u$ are not in the same doublet.}
\end{figure}
Having said that, let $Z_{f_d\bar f_df'_u\bar f'_u}^{\rm tr}$ be the
truncated Green's function associated to our four fermion process, where $f_d$ 
(respectively $f'_u$) represents a down-(respectively, up-) type lepton.
If we assume that $f'_u$ and $f_d$ are not in the same doublet, the latter four-point function
allows for the following decomposition (see Fig.\ref{Zdeco})
\bea
Z_{f_d\bar f_df'_u\bar f'_u}^{\rm tr} &=& i\Gamma_{f_d\bar f_df'_u\bar
  f'_u}-
\left(\Gamma_{f_d\bar f_d\Phi}\Delta_{\Phi\Phi'}
\Gamma_{\Phi'f'_u\bar f'_u}+\Gamma_{f_d\bar f'_u\Phi}
\Delta_{\Phi\Phi'}
\Gamma_{\Phi'f'_u\bar f_d}\right) \nonumber \\
&=&i\Gamma_{f_d\bar f_df'_u\bar
  f'_u}-\Gamma_{f_d\bar f_d\Phi}\Delta_{\Phi\Phi'}
\Gamma_{\Phi'f'_u\bar f'_u},
\label{amplidec}
\eea
where a sum over repeated fields (running over all the allowed SM 
combinations) is understood, $\Delta_{\Phi\Phi'}(q)$ indicates a (full) propagator
between the SM fields $\Phi$ and $\Phi'$, and
we have omitted the momentum dependence of the Green's functions as well as Lorentz indices.
Then, the gauge invariance of the $S$-matrix implies
\bea
\partial_{\xi_i} Z_{f_d\bar f_df'_u\bar f'_u}^{\rm tr}&=&
i\partial_{\xi_i}\Gamma_{f_d\bar f_df'_u\bar
  f'_u}\nonumber \\
&-&(\partial_{\xi_i}\Gamma_{f_d\bar f_d\Phi}) \Delta_{\Phi\Phi'}
\Gamma_{\Phi'f'_u\bar f'_u}-\Gamma_{f_d\bar f'_d\Phi}
(\partial_{\xi_i}\Delta_{\Phi\Phi'})
\Gamma_{\Phi'f'_u\bar f'_u}-\Gamma_{f_d\bar f_d\Phi}\Delta_{\Phi\Phi'}
(\partial_{\xi_i}\Gamma_{\Phi'f'_u\bar f'_u})\nonumber \\
&=&0.
\label{N1}
\eea

The NIs can be used to determine how the perturbative series rearranges itself 
in order to fulfill the above equality. 
Neglecting terms that either vanish due to the on-shell
conditions of the external fermions or cancel when using
the LSZ reduction formula, the NIs for the various terms that appear
in Eq.(\ref{N1}) can be derived from the master equation (\ref{NI}), 
and read 
\bea
\partial_{\xi_i}\Delta_{\Phi\Phi'}&=&
\Delta_{\Phi\Phi''}\Gamma_{\eta_i\Phi''{\Phi'}^{*}}+
\Gamma_{\eta_i\Phi^{*}\Phi''}\Delta_{\Phi''\Phi'},
\nonumber \\
-\partial_{\xi_i}\Gamma_{f\bar f\Phi}&=&
\Gamma_{\eta_i{\Phi'}^{*}f\bar f}\Gamma_{\Phi'\Phi}
+\Gamma_{\eta_i\Phi{\Phi'}^{*}}\Gamma_{\Phi'f\bar f},
\nonumber \\
-\partial_{\xi_i}\Gamma_{f_d\bar f_df'_u\bar
  f'_u}&=&
\Gamma_{f_d\bar f_d\Phi}
\Gamma_{\eta_i{\Phi}^{*}f'_u\bar f'_u}+\Gamma_{f_d  \bar
  f'_u\Phi}\Gamma_{\eta_i{\Phi}^{*}f'_u\bar f_d}+
\Gamma_{\eta_i{\Phi}^{*}f_d\bar f_d}\Gamma_{\Phi f'_u\bar
  f'_u}
+\Gamma_{\eta_i{\Phi}^{*}f_d\bar f'_u}\Gamma_{\Phi f'_u\bar f_d}
\nonumber \\
&=&\Gamma_{f_d\bar f_d\Phi}
\Gamma_{\eta_i{\Phi}^{*}f'_u\bar f'_u}+
\Gamma_{\eta_i{\Phi}^{*}f_d\bar f_d}\Gamma_{\Phi f'_u\bar f'_u}.
\label{NIBVP}
\eea
Notice that despite their appearance, the four point functions 
$\Gamma_{\eta_i{\Phi}^{*}f\bar f}$ (respectively, the three point functions 
$\Gamma_{\eta_i\Phi{\Phi'}^{*}}$) are vertex-like (respectively propagator-like), 
due to the static nature of the $\eta_i$ sources.

As far as the cancellations of gauge-dependent pieces are concerned, 
we see that the gauge-fixing-parameter-dependence of the internal self-energies
cancels according to the pattern
\bea
&&
\Gamma_{f_d\bar f_d\Phi}
(
\Delta_{\Phi\Phi''}\Gamma_{\eta_i\Phi''{\Phi'}^{*}}+
\Gamma_{\eta_i\Phi^{*}\Phi''}\Delta_{\Phi''\Phi'})
\Gamma_{\Phi'f'_u\bar f'_u}\nonumber \\
&-&
(\Gamma_{\eta_i\Phi{\Phi''}^{*}}
\Gamma_{\Phi''f_d\bar f'_d})
\Delta_{\Phi\Phi'}\Gamma_{\Phi'f'_u\bar f'_u}+\Gamma_{f_d\bar f_d\Phi}
\Delta_{\Phi\Phi'}
(\Gamma_{\eta_i\Phi'{\Phi''}^{*}}\Gamma_{\Phi''f_d\bar f_d})
=0.
\eea
Using finally, the relation $\Delta_{\Phi\Phi''}\Gamma_{\Phi''\Phi'}=i\delta_{\Phi\Phi'}$,
one can uncover the cancellation happening between the boxes and vertices,
according to the rules
\bea
-i\Gamma_{\Phi f_d\bar f_d}\Gamma_{\eta_i{\Phi}^{*}f_u\bar f_u}
+\Gamma_{f_d\bar f_d\Phi} \Delta_{\Phi\Phi'}\Gamma_{\Phi'\Phi''}
\Gamma_{\eta_i{\Phi''}^{*}f'_u\bar f'_u}&=&0\nonumber\\
-i\Gamma_{\eta_i{\Phi}^{*}f_d\bar f_d}\Gamma_{f'_u\bar f'_u\Phi}
+\Gamma_{\eta_i{\Phi''}^{*}f_d\bar f_d}\Gamma_{\Phi''\Phi}\Delta_{\Phi\Phi'}
\Gamma_{\Phi'f'_u\bar f'_u}&=&0.
\label{bv}
\eea

From the above patterns one concludes that the gauge
cancellations: ({\it i})  go through without the need of integration over the
virtual momenta; ({\it ii}) follow the $s$-$t$ cancellations 
characteristic of the PT (which, in a sense, is to be expected since
both the PT cancellations and the NIs are BRST-driven); ({\it iii}) 
proceed regardless of whether the fermions are massive or massless.

\subsection{The one-loop case re-examined}

In order to make the conclusions drawn at the end of the previous section more quantitative, we consider the one-loop case already addressed in the PT framework, and identify the pieces that cancel between box and vertex diagrams in the NIs.
At one-loop level, the third equation in (\ref{NIBVP}), reads
\be
-\partial_{\xi_i}\Gamma^{(1)}_{f_d\bar f_df'_u\bar f'_u}=
\Gamma^{(0)}_{f_d\bar f_d\Phi}
\Gamma^{(1)}_{\eta_i{\Phi}^{*}f'_u\bar f'_u}+
\Gamma^{(1)}_{\eta_i{\Phi}^{*}f_d\bar f_d}\Gamma^{(0)}_{\Phi f'_u\bar f'_u}.
\ee

\begin{figure}[t]
\includegraphics[width=12cm]{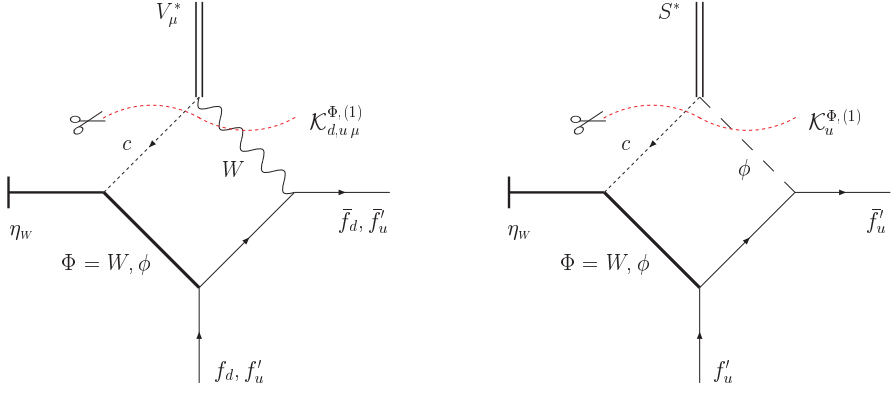}
\caption{\label{1lNI} The (one particle irreducible) diagrams contributing to the NI
of the one loop box diagrams of the $f_df'_u\to f_d f'_u$
process. Notice that since the sources $\eta_i$ are static, these
diagrams are vertex-like. The associated kernels ${\cal K}$ is also
shown, and is obtained by simply factoring out from the corresponding
diagram the vertex involving the antifield.} 
\end{figure}

We are interested in the case where $\xi_i=\xiw$ (it is immediate to show from the above equation 
and the Feynman rules provided in Fig.\ref{BRSTfr} and \ref{NIfr}, that boxes involving neutral 
gauge bosons form a gauge independent sub-set, at this order), 
so that one has
\bea
-\partial_{\xiw}\Gamma^{(1)}_{f_d\bar f_df'_u\bar f'_u}&=&
\Gamma^{(0)}_{f_d\bar f_d\Phi}
\Gamma^{(1)}_{\etaw{\Phi}^{*}f'_u\bar f'_u}+
\Gamma^{(1)}_{\etaw{\Phi}^{*}f_d\bar f_d}\Gamma^{(0)}_{\Phi f'_u\bar f'_u}\nonumber \\
&=&\Gamma^{(0)}_{f_d\bar f_d V^\mu}\Gamma^{(1)}_{\etaw V^*_\mu f'_u\bar f'_u}
+\Gamma^{(1)}_{\etaw V^{*}_\mu f_d\bar f_d}\Gamma^{(0)}_{V^\mu f'_u\bar f'_u}
+\Gamma^{(0)}_{f_d\bar f_d S}\Gamma^{(1)}_{\etaw S^* f'_u\bar f'_u}.
\label{oneloopNI}
\eea
Introducing the kernels ${\cal K}$ according to Fig.{\ref{1lNI}}, one has
\bea
\Gamma^{(0)}_{f_d\bar f_d V^\mu}\Gamma^{(1)}_{\etaw V^*_\mu f'_u\bar f'_u}&=& 
-\gw C_V\Gamma^{(0)}_{V^\mu f_d\bar f_d}\bigg[{\cal K}^{W,(1)}_{u\,\mu}+
{\cal K}^{\phi,(1)}_{u\,\mu}\bigg],
\nonumber \\
\Gamma^{(1)}_{\etaw V^{*}_\mu f_d\bar f_d}\Gamma^{(0)}_{V^\mu f'_u\bar f'_u}&=&
-\gw\left[{\cal K}^{W,(1)}_{d\,\mu}+
{\cal K}^{\phi,(1)}_{d\,\mu}\right]C_V\Gamma^{(0)}_{V^\mu f'_u\bar f'_u},
\nonumber \\
\Gamma^{(0)}_{f_d\bar f_d S}\Gamma^{(1)}_{\etaw S^* f'_u\bar f'_u}&=&
-\gw\left[{\cal K}^{W,(1)}_d+
{\cal K}^{\phi,(1)}_d\right]\left(i\gw\frac{m_{f_d}}{\Mw}P_L\right). 
\label{onelooprel}
\eea

Now, consider first of all the case in which the down-type target fermion
$f_d$ is massless and right-handedly polarized ({\it i.e.}, a pure
positive helicity state), as was done in \cite{Bernabeu:2000hf}. Then
we know that the contribution from the box diagram with two $W$s
is zero (there is no $\Gamma^{(0)}_{W_\mu f^R_d\bar f^R_d}$
vertex in this case); thus, the left-hand side
of Eq.(\ref{oneloopNI}) vanishes identically.
As far as the right-hand side is concerned, 
the third term vanishes because $m_{f_d}=0$,
the second term because the corresponding
kernels in Eq.(\ref{onelooprel}) are zero, 
while the first one due to the relation (making
explicit the sum over the neutral gauge bosons) 
\be
\sum_V C_V\Gamma^{(0)}_{f^R_d\bar f^R_d V^\mu}=0,
\ee
which is precisely the current relation employed by the PT 
\cite{Bernabeu:2000hf}. 
Of course, due to Eq.(\ref{bv}) this will be the cancellation pattern
followed by the vertex-like gauge-dependent part of the
$\Gamma^{(1)}_{A_\mu f'_u\bar f'_u}$ and $\Gamma^{(1)}_{Z_\mu f'_u\bar
  f'_u}$ vertices. 

Next, one can consider the case in which the target
fermions $f_d$ are massless, but unpolarized. 
Then, while the third term of Eq.(\ref{onelooprel}) continue to be zero, 
the second one is not, and it is needed to cancel the vertex-like gauge
dependence of the vertex $\Gamma^{(1)}_{Z_\mu f_d\bar f_d}$ which is
now present. The first term, which cancels the vertex-like
gauge-dependent part of the $\Gamma^{(1)}_{A_\mu f'_u\bar f'_u}$ and
$\Gamma^{(1)}_{Z_\mu f'_u\bar f'_u}$ vertices, is now proportional to 
\be
-\sum_V C_V\Gamma^{(0)}_{f_d\bar f_d V^\mu}=-i\frac\gw2\gamma_\mu \PrL,
\label{PT}
\ee
which is precisely the current relation (with the spinors suppressed) 
of Eq.(\ref{CI}).
Notice that the NIs show immediately that the above PT gauge cancellation 
pattern is the same regardless of whether the iso-doublet partner 
of the $f'_u$ fermions are massless (as in \cite{Bernabeu:2000hf}) or massive 
(as in Section \ref{expl}). 

Finally, we see that relaxing the hypothesis of massless down-type target
fermions does not distort the cancellations described in the previous
case: the only difference is that the third term is no longer zero,
since it is needed to cancel the gauge dependent part  
coming from the $\Gamma^{(1)}_{Sf'_u\bar f'_u}$ which is now present.

Thus we see that considering the down-type fermion to be massless or massive, 
has no bearing whatsoever 
on one's ability to implement the gauge cancellations following the PT 
algorithm.

\subsection{All-order considerations}

We next show that the pattern of gauge-cancellations which has been
unravelled at one-loop level persists to all orders in perturbation theory.
In particular, we will show that at higher orders 
the gauge-dependent terms stemming from the 
conventional vertex 
(which would naively form part of the NCR definition) 
cancel against similar vertex-like terms coming from box diagrams, 
folowing a pattern identical to that  
operating at one-loop. 
Thus, these gauge-dependent terms drop out from 
the NCR definition in a natural way, 
at any order in the perturbative expansion, 

The term we need to study is the vertex-like contribution coming from the 
$\Gamma_{f_d\bar f_d V_\mu}\Delta_{V^\mu\Phi}
(\partial_{\xiw}\Gamma_{\Phi f'_u\bar f'_u})$ piece of Eq.(\ref{N1}). After employing the 
second of Eqs.(\ref{NIBVP}), we get (discarding the propagator-like part)
\bea
-\Gamma_{f_d\bar f_d V_\mu}\Delta_{V^\mu\Phi}
(\partial_{\xiw}\Gamma_{\Phi f'_u\bar f'_u}) &=&
\Gamma_{f_d\bar f_d V_\mu}\Delta_{V^\mu\Phi}\Gamma_{\Phi\Phi'}
\Gamma_{\eta_i{\Phi'}^*f'_u\bar f'_u}\nonumber\\
&=&i\Gamma_{f_d\bar f_d V_\mu}\Gamma_{\etaw V^{\mu*}f'_u\bar f'_u}.
\label{last}
\eea

Now, from the Feynman rules of Fig.\ref{BRSTfr}, one has that
the vertex involving the neutral gauge bosons anti-field $A^*_\mu$ is connected 
to the one involving the $Z^*_\mu$ anti-field through the relation
\be
\cw\Gamma_{A^*_\mu W^\pm_\nu c^\mp}=-\sw\Gamma_{Z^*_\mu W^\pm_\nu c^\mp}.
\ee
On the other hand, since the $\etaw$ source couples to charged fields only 
(see Fig.\ref{NIfr}), the three-point functions $\Gamma_{\etaw V^*_\mu f\bar f}$ appearing 
in the NIs will be such that 
\be
\cw\Gamma_{\etaw A^*_\mu f\bar f}=-\sw\Gamma_{\etaw Z^*_\mu f\bar f}.
\ee
Therefore, introducing the kernels ${\cal K}^\Phi_u$ as the dressed version 
of the ones employed in the one-loop analysis of the previous section (see also Fig.\ref{1lNI}), 
this last equation shows that the NCR gauge-dependent terms of Eq.(\ref{last}) 
are of the form (making explicitly the sum over the neutral gauge bosons)
\be
i\sumv C_V\Gamma_{f_d\bar f_d V^\mu} {\cal K}^\Phi_\mu,
\ee
which is nothing but a direct 
higher order version of the identity of Eq.(\ref{PT}), encountered 
in the one-loop analysis. This term will be precisely cancelled by the 
box contribution $-i\Gamma_{f_d\bar f_d V^\mu}\Gamma_{\etaw V^{\mu*}f'_u\bar f'_u}$
appearing in the first formula of Eq.(\ref{bv}).

\section{\label{conc}Conclusions}

In this article we have addressed theoretical issues related to
the proper definition of the gauge-independent SM 
vertex describing the 
effective interaction between a neutrino and 
an off-shell photon, 
together with the electromagnetic form factor 
and the corresponding NCR obtained by it. 
This effective vertex has been constructed 
at one-loop in \cite{Bernabeu:2000hf}, in the framework of the PT, 
under the operational assumption that 
 the mass of the iso-doublet partner 
of the neutrino under consideration was vanishing
(except in infra-red divergent expressions). 
In the first part of the present article we have  demonstrated 
through 
an explicit calculation that 
any additional 
gauge-dependent contributions proportional to the fermion masses 
cancel partially against
similar contributions stemming from 
graphs containing would-be Goldstone bosons, and partially against 
vertex-like contributions concealed inside box-graphs. 
This cancellation proceeds precisely as the well-defined PT 
methodology dictates, without any additional assumptions
This completes the proof that the properties
of the effective NCR listed in the Introduction maintain their 
validity in the presence of massive fermions.
In the second part of the paper we have employed the powerful 
formalism of the Nielsen identities, and furnished an 
all-order demonstration of the relevant 
gauge-cancellations, 
a fact which finally allows for the all-order
definition of the effective NCR. 

\medskip 

{\it Acknowledgments:} This work has been supported by the 
CICYT Grant FPA2002-00612.


\begin{thebibliography}{99}
\bibitem{footnote}
Of course, the covariant generalization of this definition, 
just as that of a static charge distribution, is ambiguous.
\end{thebibliography}
\end{document}